# On the Influence of Weak Magnetic and Electric Fields on the Fluctuations of Ionic Electric Currents in Blood Circulation

Zakirjon Kanokov [1,2,*], Jürn W. P. Schmelzer [1,3], Avazbek K. Nasirov [1,4]

[1] Bogoliubov Laboratory of Theoretical Physics, Joint Institute for Nuclear Research, Dubna, Russia

[2] Faculty of Physics, M. Ulugbek National University of Uzbekistan, Tashkent, Uzbekistan

[3] Institut für Physik, Universität Rostock, Rostock, Germany

[4] Institute of Nuclear Physics, Tashkent, Uzbekistan

**Abstract:** An analysis of a variety of existing experimental data leads to the conclusion on the existence of a resonance mechanism allowing weak magnetic fields to affect biological processes. These fields may either be static magnetic fields comparable in magnitude with the magnetic field of the earth or weak ultra-low frequency time-dependent fields. So far, a generally accepted theoretical model allowing one to understand the effect of magnetic and electric fields on biological processes is not available. By this reason, it is not clear which characteristics of the fields, like magnetic and electric field strength, frequency of change of the field, shape of the electromagnetic wave, the duration of the magnetic or electric influence or some particular combination of them, are responsible for the biological effect. In the present analysis it is shown that external time-independent magnetic fields may cause a resonance amplification of ionic electric currents in biological tissues and, in particular, in the vasculature system due to a Brownian motion of charges. These resonance electric currents may cause necrotic changes in the tissues or blood circulation and in this way significantly affect the biological organism. The magnitude of the magnetic fields leading to resonance effects is estimated, it is shown that it depends significantly on the radius of the blood capillaries.

*E-mail address: zokirjon@yandex.ru, shokirjon2006@yandex.ru

## 1. Introduction

In recent years, the interest in a wide spectrum of noise-induced phenomena is considerably increasing [1]. This interest is to a large extent caused by a variety of experimental data proving that noise, being an inherent ingredient of any processes in real systems, may lead not only to some thermodynamic non-structural effects but, in contrast, may be the origin of macroscopically measurable phenomena or, in other words, be the origin of macroscopically observable structures..

For example, in samples, where some weak electric current is sustained, there exists the so-called Johnson noise [2], i.e., rapidly changing and unpredictable fluctuations of the current may be observed. These fluctuations are caused by molecular processes of the carriers of the charges. In systems, where the carriers of the electric charges are ions, the fluctuations in the motion of these ions yield the main contribution to the fluctuations of the currents.

The Johnson effect is observed in systems being in a thermal equilibrium state or near to such states. The existence of weak electric currents does not change the structure of the systems considerably. And since the Brownian motion of the ions is only weakly connected with fluctuations in their number, the ensemble of ions may by considered as being near to equilibrium [3]. If, in contrast, some finite electric current is established in the system, then the ensemble of ions is transformed into a non-equilibrium state. As a consequence, the spectral density of electric fluctuations increases in comparison with the respective equilibrium states. This increase is conventionally denoted as a current noise.

The measurement of the characteristics of the noise spectrum of the currents allows one to establish the basic characteristics of the mechanisms of the electronic or ionic transport. These characteristics cannot be reproduced by measurements of only the average value of the current. For example, in [4], the spectra of noise of electric currents have been reported obtained for samples of the system **n-InSb** (the concentration of free charge carriers generating the currents are equal to $n=7\cdot10^{13}$ cm$^{-3}$, $B=1$T, $T=75$K) with similar electro-physical parameters. Despite this similarity, the spectra of noise of the different samples differ not only in magnitude but also even in shape. The magnetic field changes thus both parameters the magnitude and shape of the spectra of noise. In [5] it is shown that in the temperature range 4.5<$T$<12K an intensive current noise is found in samples of **n-InSb** (here the concentration of free charge carriers is equal to $n=7\cdot10^{13}$ cm$^{-3}$). The noise measured depends non-monotonically on temperature and on the magnitude of the average current at fixed temperature.

In electrolyte solutions, diffusion may lead to additional specific effects in the generation of noise which are not observed in solid conductors. In the present paper, we investigate one particular of such example, the generation of ionic currents in water solutions in the presence of weak electric and magnetic fields. The analysis will be performed based on the Langevin equation. As it will be shown below these effects are of particular relevance for the understanding of biological processes in blood circulation and related problems as discussed below.

In the current literature, a huge amount of experimental data can be traced demonstrating the pronounced effect of ultralow-frequency or static magnetic and electric fields on biological processes. For example, fields with a magnitude of the magnetic induction of the order of 10–1000 nT may have significant biological effects [6-7] although the energy in such fields has small values as compared with the characteristic energy $k_BT$ of the chemical reactions. A generally accepted mechanism allowing one to interpret such kind of effects is not available so far. In the reviews [7-8] some theoretical models and approaches to the interpretation of these effects are summarized indicating at the same time their partly severe limitations in giving a satisfactory explanation of the effects of weak magnetic and electric fields on biological processes. The aim of the present analysis is to present a mechanism which is of very general nature and allows one to explain adequately a variety of phenomena of interaction of magnetic and electric fields with biological organisms. The basic idea is the following.

It is well-known that any biological organism represents from a thermodynamic point of view an open non-equilibrium system [9]. The functioning is realized to a huge extent by the system of blood circulation via an extended system of veins of different length and diameter. The circulation of blood is connected with the transport of electric charges. The state of these charges may change continuously due to internal processes in the living organism and the interactions with external electromagnetic fields. These fields may be generated by surrounding sources one of them being the electromagnetic field of the Earth and result in the evolution of ionic currents. Another part of the living organisms, where electric currents may occur and affect the functioning, are the living cells. They are filled and enclosed by electrolyte solutions of a variety of chemical species and compounds giving rise to the possibility of evolution of ionic electric currents.

Ionic electric currents have been shown to be of outstanding significance in the understanding of a variety of phenomena in medicine and biophysics [3, 10]. By this reason, the analysis of possible sources of evolution of such currents is of outstanding interest. Here we would like to demonstrate that the Brownian motion of charges under the influence of electric and magnetic fields can be one source of evolution of ionic electric currents in biological tissues exhibiting, at part, a pronounced resonance character.

## 2. The Langevin Equation and its Solution

### 2.1 Derivation of the Basic Equation

Stochastic methods of analysis are applied widely in different fields of physics, biophysics, biology, chemistry, and engineering [2]. For example, in [11], employing Non-Markovian Langevin equations, the properties of open quantum systems in an external magnetic field have been analyzed. In the present analysis, Langevin equations in the

Markovian limit are employed for the analysis of the effect of external magnetic and electric fields on the evolution of ionic electric currents in biological tissues. The Langevin approach gives a comprehensive description of the random walk of small particles and correspondingly allows one to account for different factors affecting the motion of the respective particles.

In the present analysis, the extensive parameters which change in the Brownian motion are the momentum, $\vec{P}_i$, of the charge carriers, $\vec{P}_i = m\vec{v}_i$, having a mass $m$ and a velocity, $\vec{v}_i$. In an equilibrium state, the linear equations determining the momentum $\vec{P}_i$ and the depending on it parameters of the system are given by

$$\frac{d\vec{P}_i}{dt} = -\xi_i \vec{v}_i + \vec{F}_e + \vec{f}_i(t) \ . \tag{1}$$

Here $\vec{F}_e$ is the external force and $\vec{f}_i(t)$ is the random force acting on the respective particle. It is the same for any atom of the same type and is not correlated for different ions

$$<\vec{f}_i(t)> = 0, \qquad <\vec{f}_k(t)\vec{f}_j(t')> = 2kT\xi_i \delta_{kj}\delta(t-t') \ . \tag{2}$$

Here $k_B$ is the Boltzmann constant and $\xi_i$ the friction coefficient.

In order to determine the density of ionic electric currents caused by the motion of positively and negatively charged ions, we have to compute the flux density of ions and to multiply it with the charge ($q$)

$$\vec{j} = q \sum_{j=1}^{n} \frac{\vec{v}_j}{V} \ . \tag{3}$$

Here the sum has to be taken over all charges of the different types of ions, which are located in a small volume, $V$, $n$ is its number. If this volume has a transverse to the ionic electric current section $S$ and a length $L$ into the direction of the current, the current is determined by the equation

$$\vec{i} = \frac{q}{L} \sum_{j=1}^{n} \vec{v}_j \ . \tag{4}$$

Employing the relation $\vec{P}_i = m\vec{v}_i$ and taking into consideration Eq. (4), from Eq.(1) we obtain the following linear stochastic differential equation

$$\frac{d\vec{i}(t)}{dt} = -\lambda \vec{i}(t) + \frac{qn}{mL}\vec{F}_e + \vec{f}(t). \tag{5}$$

The external magnetic field acts on the electric current of length $L$ with the force

$$\vec{F}_e = i\left[\vec{L}\vec{B}\right], \tag{6}$$

We will determine in the subsequent analysis fluctuations of a scalar quantity, the magnitude of the current. By a substitution of Eq.(6) into Eq.(5) we arrive at

$$\frac{di(t)}{dt} = -\Lambda i(t) + f(t), \tag{7}$$

where the notations

$$\Lambda = \lambda - \frac{qn}{m}B\sin\alpha, \qquad \lambda = \frac{\xi_j}{m}, \qquad \sin\alpha = \sin(\vec{L}\vec{B}) \tag{8}$$

are introduced. As mentioned, the random force

$$f(t) = \frac{q}{mL}\sum_j f_j(t) \tag{9}$$

is supposed to describe white noise, so we have

$$<f(t)> = 0, \qquad <f(t)f(t')> = \gamma\delta(t-t'), \tag{10}$$

where

$$\gamma = \frac{2k_B T q^2 n \xi}{m^2 L^2}. \tag{11}$$

**2.2  Solution of the Basic Equation and General Analysis**

As shown above, the electric current may be described by a linear stochastic differential equation with white noise as the random source. The respective process is a process of Ornstein-Uhlenbeck type and the formal solution of Eq.(7) may be written in the form

$$i(t) = e^{-\Lambda t}i(0) + \int_0^t e^{-\Lambda(t-\tau)} f(\tau)d\tau. \tag{12}$$

The average value of the current may be computed from Eq.(12) as

$$<i(t)> = e^{-\Lambda t} <i(0)>. \tag{13}$$

One can also easily compute the dispersion of the fluctuations, i.e. $\delta \vec{i} = \vec{i}(t) - <\vec{i}(t)>$. This quantity is given by

$$\sigma(t) = <\delta i(t)\delta i(t)> = e^{-2\Lambda t}\int_0^t e^{2\Lambda \tau}\gamma d\tau. \tag{14}$$

Taking the derivative of Eq.(14) with respect to time, we obtain

$$\frac{d\sigma(t)}{dt} = -2\Lambda\sigma(t) + \gamma. \tag{15}$$

The solution of Eq. (15) with the initial condition $\sigma(0) = 0$ has the form

$$\sigma(t) = \sigma(\infty)(1 - e^{-2\Lambda t}), \tag{16}$$

where

$$\sigma(\infty) = \lim_{t\to\infty}\sigma(t) = \frac{\gamma}{2(\lambda - \frac{qnB\sin(\alpha)}{m})}. \tag{17}$$

As evident from Eqs.(16) and (17), the fluctuations of the ionic electric current have a resonance character at

$$\lambda = \frac{qnB\sin(\alpha)}{m}. \tag{18}$$

If instead of the magnetic field, a constant electric field $E$ is applied, then Eq.(7) gets the following form

$$\frac{di(t)}{dt} = -\Lambda i(t) + \frac{\lambda EL}{R} + f(t), \tag{19}$$

where $R = \frac{mL^2\lambda}{nq^2}$ is the electric resistance. Eq.(19) has the following formal solution

$$i(t) = \left(i(0) - \frac{\lambda EL}{\Lambda R}\right)e^{-\Lambda t} + \frac{\lambda EL}{\Lambda R} + \int_0^t e^{-\Lambda(t-\tau)} f(\tau)d\tau . \tag{20}$$

For the average current, we get

$$<i(t)> = <\left(i(0) - \frac{\lambda EL}{\Lambda R}\right)> e^{-\Lambda t} + \frac{\lambda EL}{\Lambda R} . \tag{21}$$

The dispersion $\sigma(t)$ is again determined via Eq.(16).

If there is no external magnetic field, then Eq.(19) gets the following form

$$\frac{di(t)}{dt} = -\lambda i(t) + \frac{EL}{R} + f(t) . \tag{22}$$

The solution of this equation is given by

$$i(t) = \left(i(0) - \frac{EL}{R}\right)e^{-\Lambda t} + \frac{EL}{R} + \int_0^t e^{-\Lambda(t-\tau)} f(\tau)d\tau \tag{23}$$

and widely determined by the exponential term, i.e., it changes exponentially from the initial value $i(0)$ to the respective value at time $t$, $i(t)$. Since the parameter $\lambda$ is positive, in a period of time $\tau = 1/\lambda$ the average current $\bar{i}(t)$ asymptotically approaches the stationary value $i_{st} = \frac{EL}{R}$. The equations for the fluctuations of the currents with respect to this stationary value are given by

$$\frac{d\Delta i(t)}{dt} = -\lambda \Delta i(t) + f(t) . \tag{24}$$

Here $f(t)$ accounts for the effect of white noise and the process is an Ornstein-Uhlenbeck process, again. Eq.(7) can be solved consequently similarly to Eq.(7). The respective results show that in the absence of magnetic fields, resonance effects do not occur.

### 2.3. Numerical Estimates

In order to estimate the resonance value of the magnetic induction as described by Eqs. (17) and (18), we employ the following data [12-14]: the organism of human beings consist of 60-64 % water, 30-34 % organic and 6% inorganic compounds. For a person of 70 kg weight, we have an amount of 1.7 kg calcium, 0.25 kg potassium, 0.07 kg sodium, 0.042 magnesium, 0.005 kg iron, 0.003 kg zinc. The effect of calcium in the organism of a human being is very significant. Its salts are a permanent constituent of the blood, of the cell and tissue fluids. Calcium is a component part of the cell nucleus and plays a major role in the processes of cell growth. 99% of the calcium is concentrated in the bones, the remaining part in the blood system and tissues.

The blood composes about 8.6% of the mass of a human body. Hereby the fraction of the blood located in the arteries is lower than 10% of its total amount. The same amount of blood is contained in the veins, the remaining 80% are contained in smaller units like the microvasculature, arterioles, venules and capillaries. The typical values of the viscosity of blood plasma of a healthy human being at 37°C are $1.2 \cdot 10^{-3}$ Pa·s. The density of the blood is of the order $\rho = (1.06-1.064) \cdot 10^3$ kg/m$^3$. Having at one's disposal the radius of the ions, we may determine the coefficient of diffusion getting as an estimate $D = (1.8-2.0) \cdot 10^{-9}$ m$^2$/s. The value of the coefficient of friction is obtained then as $\lambda = (3-6) \cdot 10^{13}$ s$^{-1}$.

The aorta can be considered as a canal with a diameter of $(1.6–3.2) \cdot 10^{-2}$ m and a cross section area of $(2.0–3.5) \cdot 10^{-4}$ m$^2$, which splits of step by step into a network of $10^9$ capillaries each of them having a cross section area of about $7.01 \cdot 10^{-12}$ m$^2$ with an average length of about 0.1cm. The number of calcium ions in a volume $V=(2-3.5) \cdot 10^{-6}$ m$^3$ of the aorta is equal to $(0.8-1.4) \cdot 10^{19}$, in a volume $V=(7 \cdot 10^{-15})$ m$^3$ of the capillary we have $n=2.7 \cdot 10^{10}$. Substituting these values into Eq.(18), we get, at $\sin(\alpha) \approx 1$, for the aorta $B = 0.5 \cdot 10^{-12}$ T and for the capillary $B=270$ µT. These estimates show that large vasculatures are more sensitive to ultra-weak and capillaries are sensitive to weak and moderate magnetic fields.

**Conclusions**

Capillaries represent the overwhelming part of the system of blood circulation of human beings in addition to the heart, arteries, arterioles, veins and venules. In almost any of the organs and tissues of the organism these micro-vessels form networks similar to spider-webs. This whole complex system of blood circulation, including heart, vasculature, and also the mechanisms of neural and endocrinal regulation, has been created by nature in order to supply via the capillaries blood containing essential "food" required for a normal functioning of the tissues and cells. When this blood circulation is disturbed, in the cells and tissues undesired necrotic changes may occur, the cells and tissues may die or may be heavily damaged. Fluctuations of ionic electric currents may be the origin for such disturbances and may thus negatively affect these very important parts of the blood circulation system.

The financial support from the Deutsche Forschungsgemeinschaft (DFG 436 RUS 113/705/0-3) is gratefully acknowledged.